\documentclass[cameraready]{Interspeech}

\usepackage{array, amssymb, colortbl, pifont, booktabs, makecell, amsmath, graphicx, times, epsfig, multirow, xcolor}
\usepackage{tipa}  
\usepackage[T1]{fontenc} 

\definecolor{LightBlue}{rgb}{0.85,0.92,0.96}
\definecolor{custom_gray}{gray}{0.92}
\definecolor{darkgreen}{RGB}{0,130,0}
\definecolor{darkred}{RGB}{180,0,0}
\definecolor{cellcol}{gray}{0.92}

\newcommand{\blfootnote}[1]{%
  \begingroup
  \renewcommand{\thefootnote}{}%
  \footnote{#1}%
  \addtocounter{footnote}{-1}%
  \endgroup
}

\title{ArtBoost: Synthetic Articulatory Data Augmentation for Acoustic-to-Articulatory Inversion}

\ifcameraready
  \author[affiliation={1}]{Hyung Kyu}{Kim}
  \author[affiliation={2}]{Byungchan}{Hwang}
  \author[affiliation={2}]{Hak Gu}{Kim$^{\dagger,}$}

  \address{
    $^1$ Department of Imaging Science and Arts, Chung-Ang University, South Korea \\
    $^2$ Department of Metaverse Convergence, Chung-Ang University, South Korea
  }

  \email{hyung1208@cau.ac.kr, byungchan0705@cau.ac.kr, hakgukim@cau.ac.kr}
\else
  \author[affiliation={1}]{Anonymous}{Author(s)}
  \address{$^1$ Anonymous affiliation}
  \email{anonymous@domain.com}
\fi

\keywords{acoustic-to-articulatory inversion, data augmentation, audio--visual dataset}

\begin{document}
\maketitle
\ifcameraready
  \blfootnote{\textsuperscript{\textdagger}Corresponding author.}
\fi
\begin{abstract}
Recent acoustic-to-articulatory inversion (AAI) models rely on electromagnetic articulography (EMA) data, which are costly and limited in scale. To address this limitation, we propose \textit{ArtBoost}, a novel data augmentation strategy that leverages large-scale speech--mesh datasets originally developed for speech-driven 3D facial animation to improve AAI under limited EMA supervision. \textit{ArtBoost} extracts pseudo articulatory trajectories from visible facial anchors and uses them for pre-training before fine-tuning on real EMA data. Experiments show consistent improvements in PCC and RMSE. Trajectory analyses confirm that the pseudo articulatory signals reflect physically meaningful visible articulatory dynamics. Additional evaluations across different AAI architectures demonstrate stable performance gains, indicating that \textit{ArtBoost} can be integrated into diverse AAI models. These results suggest that speech--mesh data provide an effective and scalable source of articulatory supervision for AAI.
    \ifcameraready
        Project page: \url{https://cau-irislab.github.io/Interspeech26-ArtBoost/}
    \else
    \fi
\end{abstract}

\section{Introduction}
\label{sec:intro}
Acoustic-to-articulatory inversion (AAI) or speech inversion aims to estimate articulatory movements from speech signals \cite{xie2016deep, liu2015deep}.
It has long been studied to better understand speech production and to improve articulatory-aware speech modeling.
Recently, AAI has become increasingly important in various speech-related applications that connect speech signals to physical articulatory motion, enabling production-aware speech synthesis \cite{wang2024artspeech, anumanchipalli2019speech}, articulatory-based speech analysis \cite{shahrebabaki2021acoustic, parrell2018facts, illa2018low}, and speech-driven 3D facial animation \cite{kim25r_interspeech, icip2024}.

Recent advances in deep learning have led to the emergence of learning-based approaches for acoustic-to-articulatory inversion \cite{wu2015acoustic, wu2023speaker}.
These data-driven models achieve improved performance by leveraging large amounts of paired audio–articulatory data, enabling them to capture complex speech production dynamics across diverse phonetic contexts and speaking conditions. 
However, their performance is fundamentally constrained by the availability of such paired data.
Collecting large-scale audio--articulatory datasets remains inherently challenging.
Electromagnetic articulography (EMA), a primary technique for measuring articulatory movements directly from human subjects, requires specialized hardware, controlled laboratory environments, and precise sensor placement, which limits the scale and diversity of the resulting datasets.

Several approaches have been proposed to alleviate the lack of paired audio--articulatory data in AAI \cite{shahrebabaki2021acoustic}.
Recent studies have incorporated self-supervised speech representations to improve feature robustness under scarce articulatory supervision \cite{sun2022unsupervised}, while others introduce multi-stream architectures \cite{seneviratne2019multi} or auxiliary phonetic constraints \cite{xie2016deep} to enhance generalization.
More recently, inversion targets have been extended to richer modalities such as real-time MRI \cite{ramanarayanan2018analysis, lingala2016recommendations, 10889895} or ultrasound imaging \cite{zharkova2015quantifying} in an effort to capture more complete vocal tract dynamics.
Despite these developments, most approaches remain constrained by the limited scale of paired articulatory datasets. 
As most learning-based AAI models are trained on relatively small and carefully controlled corpora, coverage of articulatory variability across speakers, phonetic contexts, and speaking conditions remains insufficient, limiting the scalability of data-driven AAI approaches.

\begin{figure}[!t]
    \begin{center}
        \includegraphics[width=1.0\linewidth]{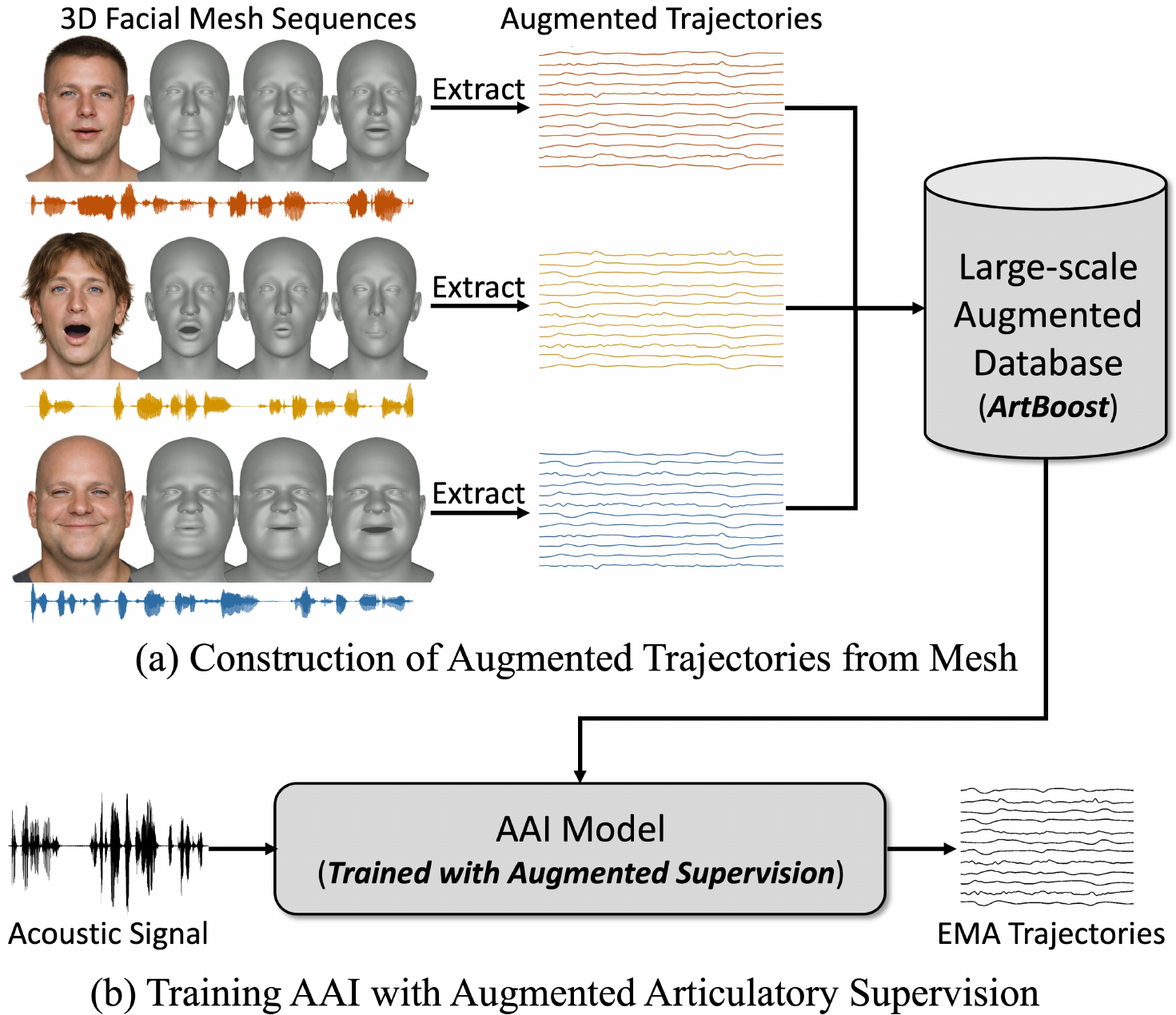}
    \end{center}
    \vspace{-15pt}
        \caption{Overview of our \textit{ArtBoost}. (a) Large-scale synthetic articulatory augmentation from audio-aligned 3D facial mesh sequences. (b) Training the AAI model using the augmented articulatory supervision.}
    \vspace{-5pt}
    \label{fig:1}
\end{figure}

We address the above limitation by introducing an augmentation strategy \textit{ArtBoost}.
Instead of modifying model architectures or relying solely on improved acoustic representations, we leverage large-scale audio-aligned 3D facial mesh sequences, which are widely used to train speech-driven 3D facial animation models (see Fig. \ref{fig:1}).
These data capture visible articulatory dynamics such as lip and jaw motion.
From this synthetic speech--mesh source, we extract articulatory motion trajectories and \textit{repurpose} them as additional supervision for AAI.
By transforming abundant synthetic speech--mesh data into articulatory motion, ArtBoost expands the effective training space without requiring additional sensor-based recordings.

We conduct comprehensive evaluations of \textit{ArtBoost} across multiple model architectures and benchmark datasets.
Summarizing the key results, applying \textit{ArtBoost} to a baseline AAI model has significantly improved the Pearson correlation coefficient (PCC) by $+2\%$ on Haskins Production Rate Comparison (HPRC) EMA dataset \cite{tiede2017quantifying} and $+25\%$ on the USC-TIMIT speech production database \cite{narayanan2014real}.

\section{Datasets} \label{sec:dataset}
\subsection{EMA Dataset} \label{sec:Datasets_1}
An EMA dataset is a paired speech-articulatory corpus used to train and evaluate AAI models, providing synchronized audio and sensor-based articulator trajectories as ground-truth.

\noindent\textbf{HPRC}~\cite{tiede2017quantifying} provides speaker-wise, utterance-level paired samples.
Each sample consists of a speech waveform of a prompted utterance and the corresponding EMA trajectories measuring articulator motion.
It includes 8 speakers and 1,440 utterances for each, forming a small but clean laboratory corpus with well-controlled recording conditions and reliable sensor trajectories. In total, this results in 11,520 utterances (7.2 hours).
For each utterance, multiple sensors attached to the speaker's articulators capture their 3D position over time, resulting in a time series of 3D trajectories for each sensor.
Audio is recorded at 44.1\,kHz and EMA is sampled at 100\,Hz.

\noindent\textbf{USC-TIMIT}~\cite{narayanan2014real} consists of utterance-level audio--EMA pairs recorded in a manner similar to HPRC, with speakers reading the same corpus used in MOCHA-TIMIT~\cite{wrench2000multichannel}.
It includes 4 speakers and 460 sentences. Overall, it contains 1,673 utterances with a total of approximately 1.2 hours.
Audio is recorded at 44.1\,kHz and EMA is sampled at 100\,Hz.

\indent Because these datasets are limited in scale and diversity due to costly and complex sensor-based data collection, we use them as the target dataset for fine-tuning and evaluation under the existing AAI preprocessing protocol~\cite{hao2024exploring}.

\subsection{Speech--Mesh Dataset} \label{sec:Datasets_2}
Large-scale speech--mesh datasets reconstructed from in-the-wild talking head videos provide temporally aligned speech signals and 3D facial mesh sequences. 
Such datasets have been widely used for speech-driven 3D facial animation~\cite{VOCA2019,meshtalk2021,faceformer2022, codetalker2023, Kim_2025_ICCV} and talking head generation~\cite{chu2024gagavatar,10657656}.

\noindent\textbf{TFHP}~\cite{diffposetalk2024} is a large-scale speech--mesh corpus reconstructed from in the wild RGB videos.
It provides paired speech and time-aligned 3D facial mesh sequences.
Audio is sampled at 16\,kHz, and meshes are provided at 25\,fps with a FLAME topology~\cite{FLAME2017} consisting of $5{,}023$ vertices.
It includes 588 subjects and 27.1 hours of paired audio--mesh data.
Unlike EMA datasets that are organized as utterance-level samples, TFHP consists of long, continuous video-level recordings per subject.

In this work, we leverage the speech--mesh dataset as a scalable source of pseudo articulatory supervision to mitigate the limited size of sensor-based EMA dataset. 
Although it does not provide direct articulatory measurements, its temporally aligned speech and mesh sequences enable the extraction of visible articulatory motion signals. 
To ensure compatibility with conventional AAI training protocols, we reorganize the long video-level recordings into utterance-level segments.

\section{Proposed Method} \label{sec:method}
\begin{figure}[t!]
    \centering
    \includegraphics[width=1\linewidth]{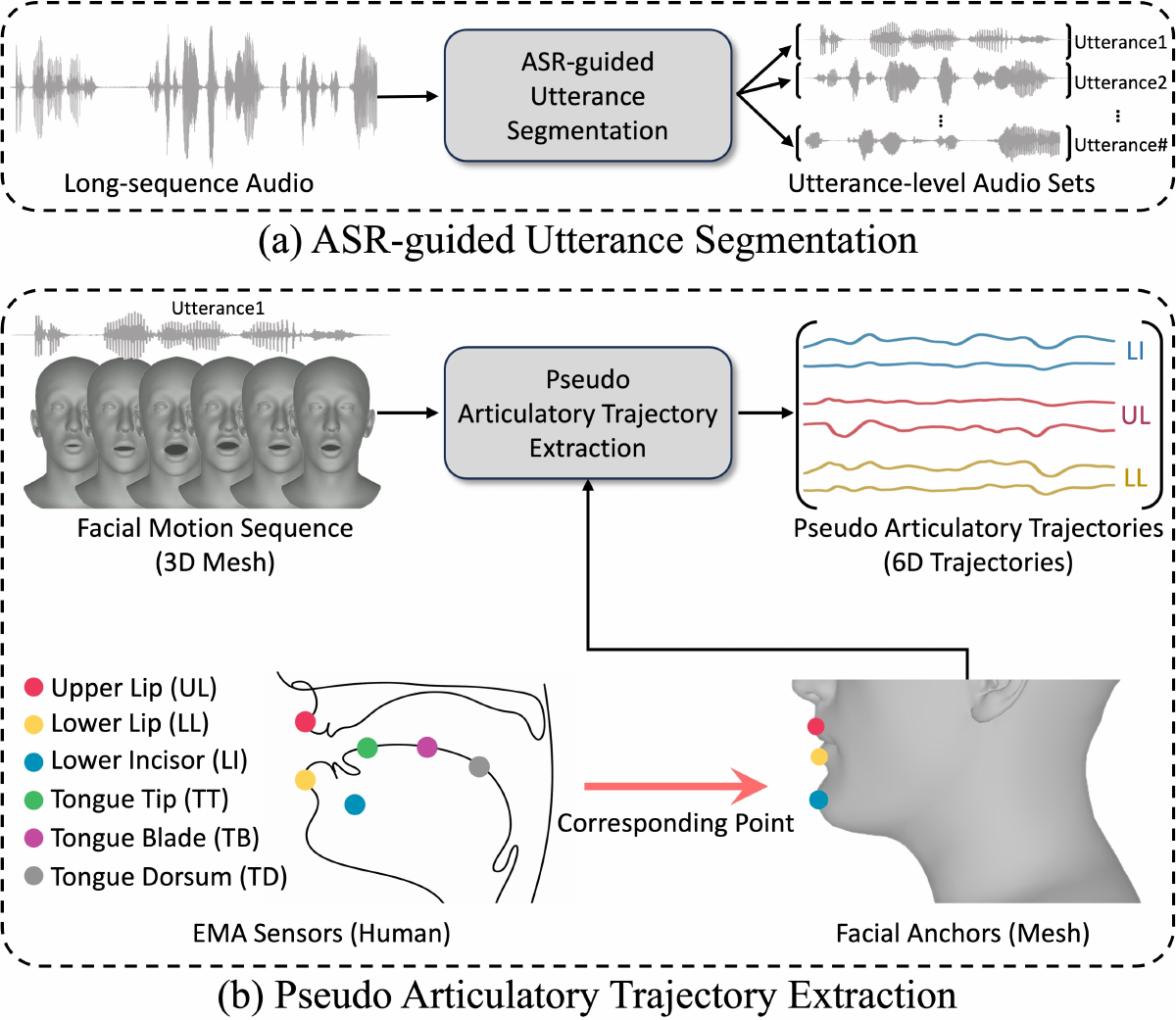}
    \vspace{-15pt}
    \caption{
        Illustration of our \textit{ArtBoost}. ASR-guided segmentation produces utterance-level clips, facial anchors are tracked to obtain articulatory trajectories, and the pseudo labels are used to pre-train the AAI model prior to EMA adaptation.
    }
    \vspace{-10pt}
    \label{fig:pipeline}
\end{figure}

Fig.~\ref{fig:pipeline} illustrates the overall pipeline of \textit{ArtBoost} for constructing large-scale pseudo articulatory supervision from speech--mesh data. 
Our goal is to alleviate the scarcity of paired audio--EMA corpora by repurposing speech--mesh recordings as an alternative source of articulatory cues.
Unlike EMA datasets, which provide direct sensor-based articulator trajectories, speech--mesh datasets contain temporally aligned audio and 3D facial mesh sequences without explicit articulatory labels. 
However, visible articulators such as the lips and jaw can be encoded in the mesh geometry and evolve consistently with speech.
We exploit this property to extract articulator-related motion signals and convert them into compatible pseudo trajectories through three steps:
(1) segmenting long video-level speech--mesh recordings into utterance-level clips to match conventional EMA settings (Sec.~\ref{sec:seg}), (2) tracking articulator-aligned facial anchors to obtain pseudo articulatory trajectories (Sec.~\ref{sec:map}), (3) pre-training the AAI model with pseudo supervision followed by fine-tuning on real EMA datasets (Sec.~\ref{sec:train}).

\subsection{ASR-guided Utterance Segmentation} \label{sec:seg}
Speech--mesh datasets consist of long video-level recordings. On the other hand, EMA corpora are organized at the utterance level.
To ensure compatibility with AAI training protocols, we segment each long recording into utterance-level clips using automatic speech recognition (ASR) (Fig.~\ref{fig:pipeline} (a)).
Given a speech signal $\mathbf{x}$ sampled at $f_s$ Hz, the ASR produces a sequence of recognized words with timestamps, $\mathcal{W}=\{(w_i,\tau_i^{s},\tau_i^{e})\}_{i=1}^{M}$ where $w_i$ denotes $i$-th word and $\tau_i^{s},\tau_i^{e}$ represent its start/end times, respectively.
We group consecutive words into $U$ utterance candidates when either (1) the inter-word silence exceeds a threshold $\Delta$, (\textit{i.e.}, $\tau_i^{s}-\tau_{i-1}^{e}>\Delta$), or (2) the current group reaches a predefined maximum number of words $K$.
For each utterance candidate spanning words $w_{i},\dots,w_{j}$, we define its temporal boundary as the interval from the first word onset $\tau_{i}^{s}$ to the last word offset $\tau_{j}^{e}$, extended by a margin of 0.1~s on both sides.
These time intervals are then converted into corresponding speech sample indices and mesh frame indices using $f_s$ and mesh frame rate $f_m$.
As a result, each long speech--mesh recording is transformed into a set of synchronized utterance-level speech--mesh pairs: $\{(\mathbf{x}^{(u)}, \mathbf{v}^{(u)})\}_{u=1}^{U}$ which are compatible with the utterance-level structure of EMA corpora.

\subsection{Pseudo Articulatory Trajectory Extraction} \label{sec:map}
Given an utterance-level mesh clip $\mathbf{v}^{(u)}$, pseudo articulatory trajectories are extracted by tracking visible facial anchors corresponding to articulators, which are the upper lip (UL), lower lip (LL), and lower incisor (LI) on the mesh surface (Fig.~\ref{fig:pipeline} (b)).
Since speech--mesh data do not provide direct articulatory measurements, we use the tracked anchors to approximate visible articulator motion.
To reduce local mesh noise and improve stability, each anchor is represented by the mean position of a predefined vertex region.
Specifically, the mean position of the UL anchor region at frame $t$ can be defined as

\begin{equation}
    \mathbf{p}^{(u)}_{\mathrm{UL},t}
    =
    \frac{1}{|\Omega_{\mathrm{UL}}|}
    \sum_{v\in\Omega_{\mathrm{UL}}}\mathbf{v}^{(u)}_t[v],
\label{eq:anchor_pos_ul}
\end{equation}
where $\Omega_{\mathrm{UL}}$ is the vertex index set of the UL anchor region and $\mathbf{v}^{(u)}_t[v]\in\mathbb{R}^{3}$ is the 3D coordinate of vertex $v$ at frame $t$.

To align with the conventional EMA trajectory representation used in AAI~\cite{wu2023speaker,hao2024exploring}, we retain motion components along the protrusion ($z$-axis) and mouth opening ($y$-axis) directions from each anchor position.

\begin{equation}
    \tilde{\mathbf{t}}^{(u)}_{\mathrm{UL},t} =
    \begin{bmatrix}
        \mathbf{p}^{(u)}_{\mathrm{UL},t}(z)\\
        \mathbf{p}^{(u)}_{\mathrm{UL},t}(y)
    \end{bmatrix}
    \in\mathbb{R}^{2}.
\label{eq:zy_map_ul}
\end{equation}

The resulting trajectories, $\tilde{\mathbf{t}}^{(u)}_{\mathrm{UL}}, \tilde{\mathbf{t}}^{(u)}_{\mathrm{LL}},$ and $\tilde{\mathbf{t}}^{(u)}_{\mathrm{LI}}$, are integrated into a 12-channel target representation $\mathbf{t}^{(u)}\in\mathbb{R}^{T_u\times12}$, following the EMA trajectory format commonly used in prior works~\cite{wu2023speaker,hao2024exploring}.
Since our pseudo articulatory trajectories are available only for the visible anchors (UL/LL/LI), we assign values to their corresponding channels and set the remaining channels to zero.
Then, the pseudo articulatory trajectories are resampled to the target articulatory frame rate via channel-wise interpolation.
Finally, the pseudo articulatory target trajectories can be obtained, $\tilde{\mathbf{t}}^{(u)}_{\mathrm{ArtBoost}}=\{\tilde{\mathbf{t}}^{(u)}_{\mathrm{ArtBoost},t}\}_{t=1}^{T_u}$.

\subsection{Training Strategy} \label{sec:train}
We train the AAI model in two steps using different sources of supervision.
First, we pre-train the model using the pseudo articulatory target trajectories $\tilde{\mathbf{t}}^{(u)}_{\mathrm{ArtBoost}}$ extracted from the speech--mesh data.
Since pseudo supervision is available only for the visible anchors (UL/LL/LI), we compute a channel-masked prediction error as follows.
\begin{equation}
    \mathcal{L}_{\mathrm{ArtBoost}} =
    \frac{1}{T_u}
    \sum_{t=1}^{T_u}
    \left\lVert
    \mathbf{m}\odot\left(\hat{\mathbf{t}}^{(u)}_t-\tilde{\mathbf{t}}_{\mathrm{ArtBoost},t}^{(u)}\right)
    \right\rVert_2^{2},
\label{eq:loss_aug_masked}
\end{equation}
where $\mathbf{m}\in\{0,1\}^{12}$ is a fixed channel mask that selects the UL/LL/LI channels and $\odot$ denotes element-wise multiplication.
$\hat{\mathbf{t}}^{(u)}_t$ is the model prediction from audio. 

After the pre-training, we fine-tune the model on real EMA trajectories $\mathbf{t}^{(u)}_t$ (ground-truth) with full-channel supervision.
\begin{equation}
    \mathcal{L}_{\mathrm{EMA}}
    =
    \frac{1}{T_u}
    \sum_{t=1}^{T_u}
    \left\lVert
    \hat{\mathbf{t}}^{(u)}_t - \mathbf{t}^{(u)}_t
    \right\rVert_2^{2}.
\label{eq:loss_ema_full}
\end{equation}
This strategy allows the model to first learn a strong prior of visible articulatory motion from large-scale pseudo supervision and then refine it using ground-truth EMA trajectories.

\newcommand{\yes}{\textbf{\scalebox{1.15}{\textcolor{darkgreen}{\ding{51}}}}}
\newcommand{\no}{\textbf{\scalebox{1.15}{\textcolor{darkred}{\ding{55}}}}}
\begin{table}[t!]
\centering
\small
\setlength{\tabcolsep}{1.5pt}
\renewcommand{\arraystretch}{1.05}
\caption{Leave-one-speaker-out results on HPRC and USC-TIMIT (mean$\pm$std).}
\label{tab:pcc_rmse_haskins_usc_with_without_tfhp}
\vspace{-8pt}
\hspace*{-1mm}
\begin{tabular}{cc|c|c|c|c}
\toprule
\multirow{3}{*}{Dataset}
& \multicolumn{1}{c|}{\multirow{3}{*}{\makecell[c]{Unseen\\[-0.1ex]Speaker}}}
& \multicolumn{2}{c|}{PCC ($\uparrow$)}
& \multicolumn{2}{c}{RMSE ($\downarrow$)} \\
\cline{3-4}\cline{5-6}
& \multicolumn{1}{c|}{}   
& \multicolumn{2}{c|}{Data Augmentation}
& \multicolumn{2}{c}{Data Augmentation} \\
& \multicolumn{1}{c|}{}   
& \no & \yes
& \no & \yes \\
\midrule

\multirow{9}{*}{HPRC}
& F01     & 0.705 & \textbf{0.720} & 0.712 & \textbf{0.695} \\
& F02     & 0.698 & \textbf{0.712} & 0.726 & \textbf{0.708} \\
& F03     & 0.640 & \textbf{0.678} & 0.776 & \textbf{0.742} \\
& F04     & 0.768 & \textbf{0.771} & 0.639 & \textbf{0.636} \\
\cmidrule(lr){2-6}
& M01     & 0.704 & \textbf{0.719} & 0.710 & \textbf{0.693} \\
& M02     & 0.633 & \textbf{0.653} & 0.777 & \textbf{0.762} \\
& M03     & 0.659 & \textbf{0.684} & 0.757 & \textbf{0.734} \\
& M04     & 0.620 & \textbf{0.647} & 0.790 & \textbf{0.765} \\
\cmidrule(lr){2-6}
& Overall & 0.678\textsuperscript{±0.05} & \textbf{0.698\textsuperscript{±0.04}}
          & 0.736\textsuperscript{±0.05} & \textbf{0.717\textsuperscript{±0.04}} \\
\cmidrule(lr){1-6}
\multirow{5}{*}{USC-TIMIT}
& F1      & 0.225 & \textbf{0.480} & 0.923 & \textbf{0.814} \\
& F5      & 0.477 & \textbf{0.585} & 0.795 & \textbf{0.738} \\
\cmidrule(lr){2-6}
& M1      & 0.278 & \textbf{0.497} & 0.907 & \textbf{0.808} \\
& M3      & 0.424 & \textbf{0.479} & 0.832 & \textbf{0.809} \\
\cmidrule(lr){2-6}
& Overall & 0.351\textsuperscript{±0.10} & \textbf{0.510\textsuperscript{±0.04}}
          & 0.864\textsuperscript{±0.05} & \textbf{0.792\textsuperscript{±0.03}} \\
\bottomrule
\end{tabular}
\vspace{-10pt}
\end{table}

\section{Experiments}
\label{sec:Experiments}

\subsection{Experimental Settings}
\subsubsection{Datasets}
In our experiments, we employ TFHP~\cite{diffposetalk2024} to generate a pseudo articulatory target trajectory dataset using our \textit{ArtBoost} for pre-training and HPRC~\cite{tiede2017quantifying} and USC-TIMIT~\cite{narayanan2014real} are employed as real EMA datasets for fine-tuning.
For fair comparison, we standardize all inputs to a common audio sampling rate and articulatory frame rate using the same preprocessing protocol and parameter settings as prior AAI work~\cite{hao2024exploring}.

\subsubsection{Implementation Details}
All experiments were conducted on a single NVIDIA RTX 3090 GPU.
We follow the input features, preprocessing protocol, and parameter settings of prior AAI work~\cite{hao2024exploring}, and use the model-specific architectures and hyperparameters reported in~\cite{wu2023speaker,hao2024exploring}.
For pre-training, we compute pseudo articulatory trajectories from FLAME-topology meshes using manually selected anchor regions. 
The UL and LL anchors are chosen to spatially correspond to the EMA sensor locations in the HPRC dataset. 
For the LI channel, we use a mesh anchor located on the lower incisor region to approximate jaw motion~\cite{lu2009experiments}.
In TFHP, long recordings are segmented via ASR~\cite{radford2023robust} (up to 7 words per utterance), and the pseudo articulatory trajectories are resampled to the target rate using cubic interpolation~\cite{fritsch1980monotone}.

\subsubsection{Evaluation Metrics}
To evaluate the effectiveness of our data augmentation strategy in AAI, we use Pearson correlation coefficient (PCC) and root mean square error (RMSE) as in ~\cite{hao2024exploring}.
PCC measures linear agreement between predicted and ground-truth trajectories.
RMSE quantifies the average magnitude of prediction error.

\begin{figure}[tp!]
\centering
\includegraphics[width=0.95\linewidth]{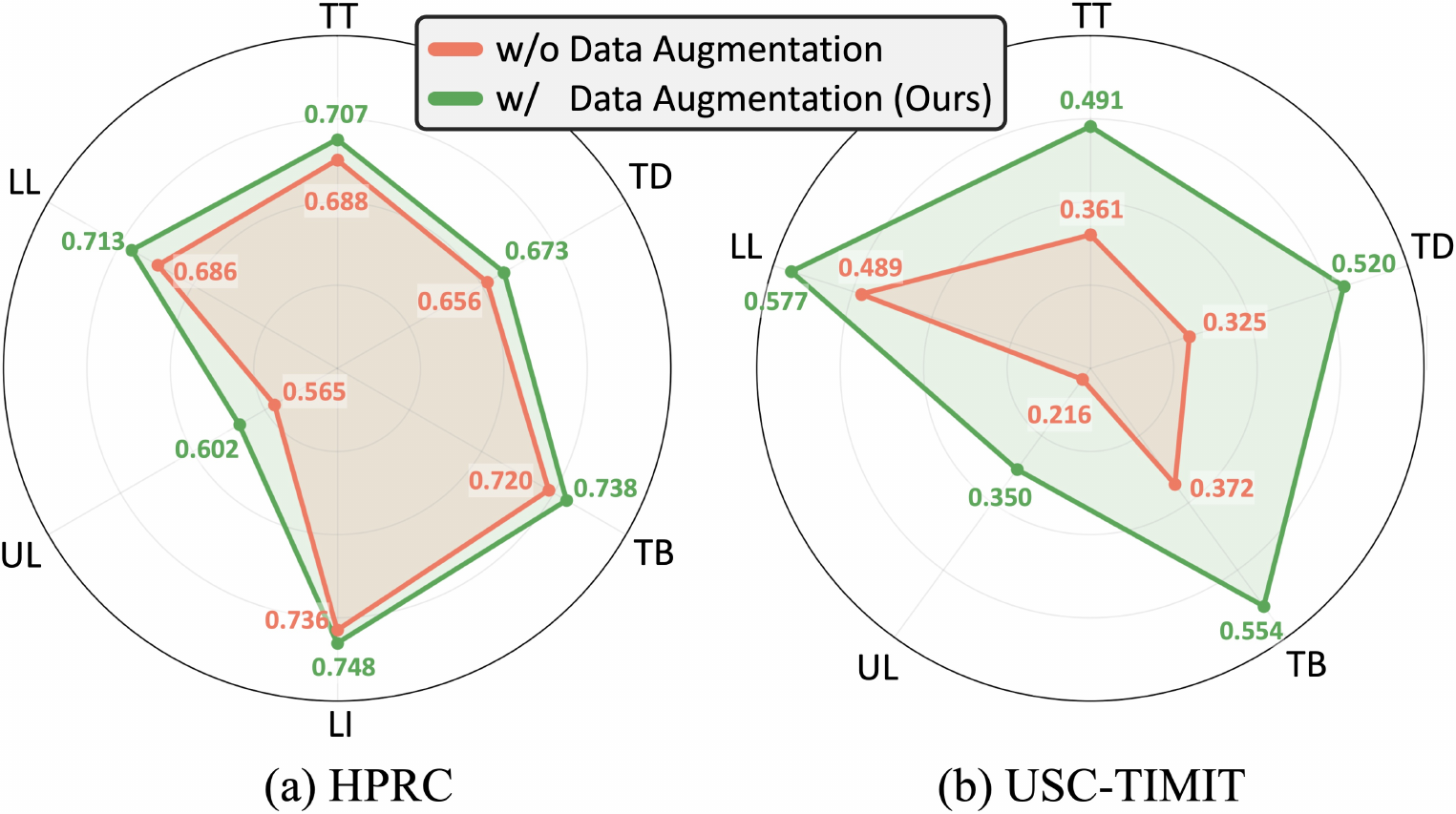}
\vspace{-5pt}
\caption
{
Articulator-wise PCC across EMA trajectories.
}
\label{fig:3}
\end{figure}

\begin{figure}[!t]
\begin{center}
\includegraphics[width=1.0\linewidth]{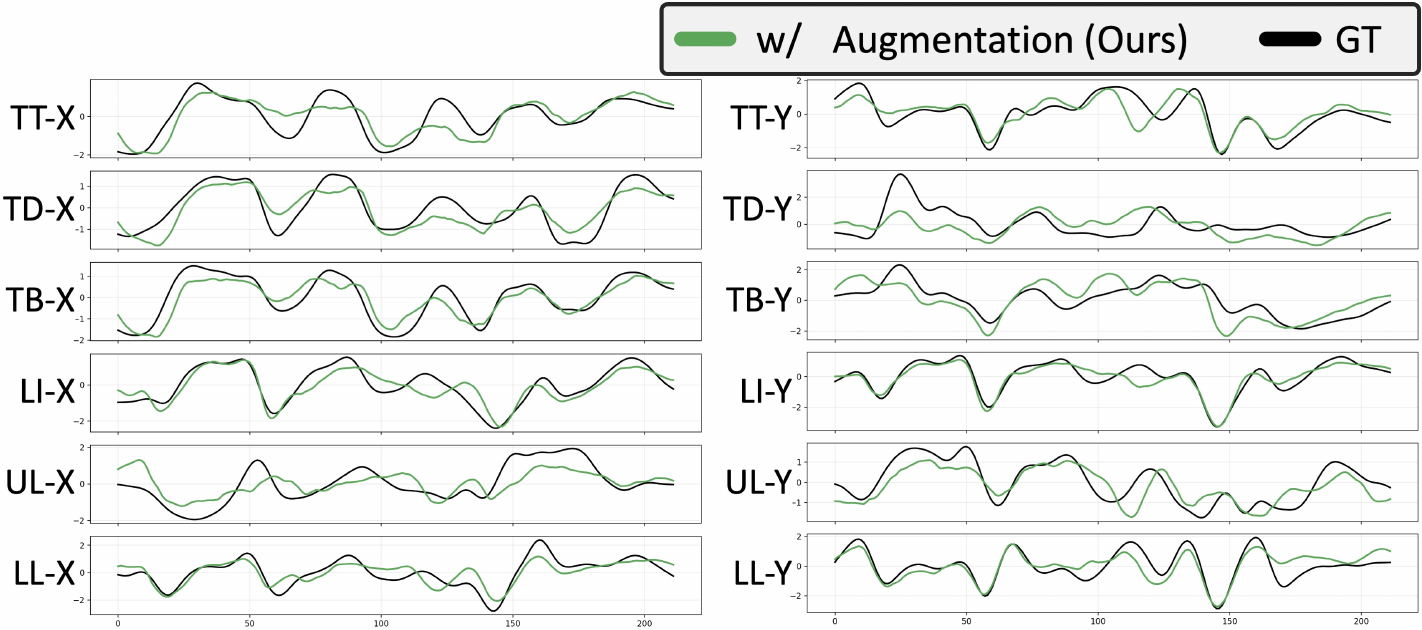}
\end{center}
\vspace{-15pt}
\caption
{
    Qualitative EMA trajectory comparison on HPRC.
}
\label{fig:4}
\vspace{-5pt}
\end{figure}

\subsection{Quantitative Results}
Table ~\ref{tab:pcc_rmse_haskins_usc_with_without_tfhp} presents leave-one-speaker-out results comparing models trained with and without our \textit{ArtBoost} augmentation.
Across both HPRC and USC-TIMIT datasets, incorporating our pseudo articulatory trajectories improves performance (i.e., higher PCC and lower RMSE).
The gains are particularly pronounced on USC-TIMIT, where the available ground-truth EMA data is smaller.
This indicates that our \textit{ArtBoost} is more effective when the amount of ground-truth EMA data is limited.

Fig.~\ref{fig:3} provides articulator-wise PCC comparisons.
Although pseudo articulatory trajectories are constructed only for visible anchors (UL/LL/LI), our \textit{ArtBoost} improves prediction performance across multiple articulators.
This means that our augmentation enhances the learned articulatory representation beyond the directly supervised channels.

\subsection{Articulatory Trajectory Comparison}
In Fig.~\ref{fig:4}, we compare predicted and ground-truth EMA trajectories on HPRC.
Across the protrusion (`-X') and aperture (`-Y') channels for multiple articulators, the predicted trajectories follow the overall temporal trends of the ground-truth signals, capturing peak movements and transition patterns.
The visualization illustrates that the model produces temporally coherent and physically plausible articulatory motions consistent with expected speech dynamics.

\begin{figure}[t]
\begin{center}
\includegraphics[width=0.95\linewidth]{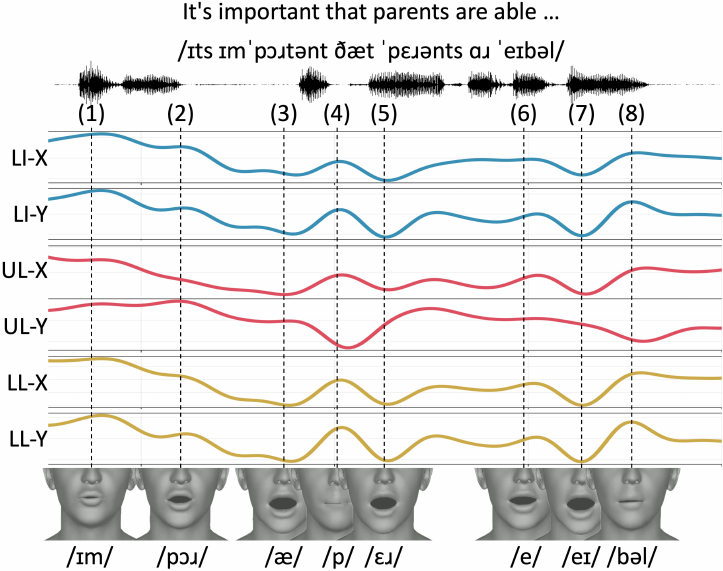}
\end{center}
\vspace{-15pt}
\caption
{
Visualization of pseudo trajectories (LI/UL/LL) and the corresponding facial mesh renderings.
}
\vspace{-5pt}
\label{fig:6}
\end{figure}

Fig.~\ref{fig:6} illustrates how pseudo articulatory trajectories are derived from speech--mesh data and how they correspond to visible facial motion.
The figure shows synchronized audio, pseudo LI/UL/LL trajectory signals, and the 3D facial mesh frames from TFHP dataset.
During bilabial closure, reduced lip opening is reflected in the Y-direction components, while protrusion-related motion appears in the X-direction.
The temporal alignment between mesh deformation and trajectory variation demonstrates that the extracted pseudo articulatory trajectories consistently reflect meaningful visible articulatory dynamics.
This visualization validates that the mesh-derived signals encode physically interpretable articulatory movements, supporting their use as training targets in the absence of direct EMA measurements.

\subsection{Results Across Different AAI Architectures}
Table~\ref{tab:2} presents results for self-supervised learning (SSL)-AAI model~\cite{hao2024exploring} and speaker-independent (SI)-AAI model ~\cite{wu2023speaker} under unseen-speaker evaluation.
For both architectures and across both datasets, incorporating our \textit{ArtBoost} leads to consistent increases in PCC and reductions in RMSE.
The performance gains observed on two structurally different AAI models indicate that the benefit of \textit{ArtBoost} is not limited to a specific architecture, but remains stable across different models.

\begin{table}[t!]
\centering
\small
\setlength{\tabcolsep}{1.5pt}
\renewcommand{\arraystretch}{1.05}
\caption{Results on HPRC and USC-TIMIT under unseen-speaker evaluation (mean$\pm$std) for different AAI models.}
\vspace{-8pt}
\label{tab:2}
\begin{tabular}{c c|c|c|c|c}
\toprule
\multirow{3}{*}{Models} & \multirow{3}{*}{Dataset}
& \multicolumn{2}{c|}{PCC ($\uparrow$)}
& \multicolumn{2}{c}{RMSE ($\downarrow$)} \\
\cline{3-4}\cline{5-6}
& & \multicolumn{2}{c|}{Data Augmentation}
  & \multicolumn{2}{c}{Data Augmentation} \\
& & \no & \yes & \no & \yes \\
\midrule
\multirow{2}{*}{\raisebox{-2.ex}{\makecell[c]{\fontsize{8.5}{8}\selectfont SSL-AAI\\\cite{hao2024exploring}}}}
& {\fontsize{8.5}{8}\selectfont HPRC}     & 0.678\textsuperscript{±0.05} & \textbf{0.698\textsuperscript{±0.04}}
          & 0.736\textsuperscript{±0.05} & \textbf{0.717\textsuperscript{±0.04}} \\
\cmidrule(lr){2-6}
& {\fontsize{8.5}{8}\selectfont USC-TIMIT} & 0.351\textsuperscript{±0.10} & \textbf{0.510\textsuperscript{±0.04}}
           & 0.864\textsuperscript{±0.05} & \textbf{0.792\textsuperscript{±0.04}} \\
\cmidrule(lr){1-6}
\multirow{2}{*}{\raisebox{-2.ex}{\makecell[c]{\fontsize{8.5}{8}\selectfont SI-AAI\\\cite{wu2023speaker}}}}
& {\fontsize{8.5}{8}\selectfont HPRC}     & 0.717\textsuperscript{±0.04} & \textbf{0.732\textsuperscript{±0.04}}
        & 0.706\textsuperscript{±0.05} & \textbf{0.689\textsuperscript{±0.04}} \\
\cmidrule(lr){2-6}
& {\fontsize{8.5}{8}\selectfont USC-TIMIT} & 0.488\textsuperscript{±0.02} & \textbf{0.593\textsuperscript{±0.03}}
         & 0.917\textsuperscript{±0.02} & \textbf{0.817\textsuperscript{±0.02}} \\
\bottomrule
\end{tabular}
\vspace{-10pt}
\end{table}

\section{Conclusion}
\label{sec:Conclusion}
We introduced \textit{ArtBoost}, a novel data augmentation strategy that leverages speech--mesh data to improve acoustic-to-articulatory inversion under limited EMA datasets. 
By extracting pseudo articulatory trajectories from visible facial anchors, \textit{ArtBoost} enables pre-training followed by fine-tuning on real EMA data.
Experiments on HPRC and USC-TIMIT demonstrate consistent improvements in PCC and RMSE. 
Trajectory-level visualizations further confirm that the extracted pseudo articulatory signals reflect physically interpretable visible articulatory dynamics. 
Additional experiments across different AAI models show that the performance gains remain consistent, indicating that \textit{ArtBoost} can be integrated into diverse AAI modeling frameworks.
These findings highlight speech--mesh data as a practical source of scalable articulatory supervision.

\section{Generative AI Use Disclosure}
During the preparation of this manuscript, the authors used OpenAI's GPT as a writing assistant for editing and language polishing. This tool was used solely to improve the readability and overall clarity of the prose. For implementation-level assistance, including code refinement and debugging, the authors utilized Anthropic's Claude. These tools were not involved in the design of the proposed methods, experimental protocols, or scientific interpretations. All content, code, and conclusions were reviewed, revised, and approved by the authors, who take full responsibility for the integrity of the manuscript and its submission.



\label{sec:refs}
\bibliographystyle{IEEEtran}
\bibliography{ref}

\end{document}